\documentclass[aps,prl,reprint,superscriptaddress,floatfix, citeautoscript]{revtex4-1}

\usepackage{amsmath}
\usepackage{color, soul}
\usepackage{graphicx}
\usepackage{epstopdf}

\definecolor{agreen}{rgb}{0, 0.44, 0}
\definecolor{indigo}{rgb}{0.58,0.34,0.92}

\newcommand{\beq}{\begin{equation}}
\newcommand{\eeq}{\end{equation}}
\newcommand{\beqy}{\begin{eqnarray}}
\newcommand{\eeqy}{\end{eqnarray}}
\newcommand{\beqs}{\begin{equation*}}
\newcommand{\eeqs}{\end{equation*}}
\newcommand{\bit}{\begin{itemize}}
\newcommand{\eit}{\end{itemize}}

\begin{document}

\title{Strain and ferroelectric soft mode induced superconductivity in strontium titanate}

\author{K. Dunnett}
\affiliation{Nordita, KTH Royal Institute of Technology and Stockholm University, Roslagstullsbacken 23, SE-106 91 Stockholm, Sweden}

\author{Awadhesh Narayan}
\affiliation{Materials Theory, ETH Zurich, Wolfgang-Pauli-Strasse 27, CH-8093 Z\"{u}rich, Switzerland}
\author{N. A. Spaldin}
\affiliation{Materials Theory, ETH Zurich, Wolfgang-Pauli-Strasse 27, CH-8093 Z\"{u}rich, Switzerland}

\author{A. V. Balatsky}
\affiliation{Nordita, KTH Royal Institute of Technology and Stockholm University, Roslagstullsbacken 23, SE-106 91 Stockholm, Sweden}
\affiliation{Institute for Materials Science, Los Alamos National Laboratory, Los Alamos, New Mexico 87545, USA}
\affiliation{Department of Physics, University of Connecticut, Storrs, CT 06269, USA}
\date{\today}

\begin{abstract}
We investigate the effects of strain on superconductivity with particular reference to SrTiO$_3$. Assuming that a ferroelectric mode that softens under tensile strain is responsible for the coupling, an increase in the critical temperature and range of carrier densities for superconductivity is predicted, while the peak of the superconducting dome shifts towards lower carrier densities. Using a Ginzburg-Landau approach in 2D, we find a linear dependence of the critical temperature on strain: if the couplings between the order parameter and strains in different directions differ while their sum is fixed, different behaviours under uniaxial and biaxial (uniform) strain can be understood.
\end{abstract}

\maketitle

Strain is one of several mechanisms by which the incipient ferroelectric \cite{PhysRevLett.19.1176, JPhysCSolStat.5.2711, PhysRevB.13.271, PhysRevB.19.3593, PhysRevLett.104.197601} strontium titanate - SrTiO$_3$ (STO) - can be made ferroelectric \cite{PhysRevB.13.271, PhysRevB.61.825, SolStatComm.9.191}. The interplay between ferroelectricity and superconductivity in STO has been investigated in the context of strontium \cite{NatPhys.13.7.643} and oxygen isotope \cite{PhysRevLett.115.247002, SciRep.6.37582} substituted STO, finding an increase in the superconducting critical temperature ($T_c$) in samples moved closer to the ferroelectric quantum critical point. Early experimental data showed that compression generally significantly reduces $T_c$ in STO, with the exception of uniaxial stress at low carrier concentrations where the critical temperature was seen to increase \cite{JLowTPhys.2.333}, but the possible connection between the ferroelectric quantum critical point and the changes in $T_c$ was not considered.

Working within the framework of ferroelectric induced superconductivity in STO \cite{PhysRevLett.115.247002}, we consider the effects of strain on the superconducting dome of STO. Based on this model we predict 1) an increase in $T_c$ under tensile strain, accompanied by a broadening of the range of carrier densities with reasonable critical temperatures, particularly at the low carrier density end, 2) a shift in the location of the peak of the superconducting dome, and 3) a sharp peak in $T_c$ signalling the limit of our model for the doping range where the ferroelectric quantum critical point is expected to occur at carrier concentrations relevant for superconductivity and small tensile strain. Although at the breakdown point of the model, we still expect the second peak to be present and have observable consequences.

We also find that, under biaxial strain, the different ferroelectric modes behave differently and the changes in $T_c$ may indicate which ferroelectric mode is most important for superconductivity. Using a simple Ginzburg-Landau model of a uniform superconducting order parameter coupled directly to applied strain, we find a linear dependence of the superconducting critical temperature on strain and go some way towards quantifying the very strong dependence of $T_c$ on strain in STO compared to elemental superconductors \cite{JLowTPhys.2.333, ContempPhys.10.4.355}. We find that the increase in $T_c$ observed in some samples under uniaxial compression \cite{JLowTPhys.2.333} can be understood qualitatively if the couplings between the superconducting order parameter and strains in different directions have different strengths.


One of the key features of superconductivity in STO is the presence of a superconducting dome where $T_c$ varies with carrier density $n$ with a maximum at `optimum' doping \cite{PhysRev.163.380, PhysRevLett.112.207002}. The generic functional dependence of the critical temperature on strain and carrier concentration $n$ can be written as $T_c = T_0f(n(u),u)$ where $T_0$ is the overall scale and $f$ being dimensionless function assumed to bounded form above at unity. Since the carrier density $n$ will be affected by changes in the volume of the unit cell. We expect that $\partial_u T_c/T_0 = f^\prime (n(u),u) = \partial_n f\partial_u n + \partial_uf$, but the changes in $n$ due to changes in the volume of the unit cell are small at small strains so $\partial_u n \approx 0$ (see supplemental material \S I).

To be specific, we use the model of Ref. \onlinecite{PhysRevLett.115.247002} (also supplemental material \S II) where ferroelectric phonons are assumed to be responsible for the superconducting pairing. This model provides a good description of the superconducting dome within a strong coupling framework \cite{PhysRevLett.115.247002} despite the fact that STO is not within the Migdal limit \cite{ArXiv:1703.04716}. In the Eliashberg strong coupling formalism, the BCS coupling constant is \cite{PhysRevLett.115.247002, PhysRevB.84.205111}:
\beq
\lambda = \int_0^\infty d\omega_q \frac{\alpha^2(\omega_q)}{\omega_q}F(\omega_q) \label{eq:lambdaInt}
\eeq
where $\omega_q$ is any phonon dispersion and $\alpha(\omega_q)$ is the electron-phonon coupling. The main features of soft mode superconductivity are captured by considering a van Hove singularity at $q=0$, for which $F(\omega_q) \sim \delta(\omega_q-\omega_0)$ so $\lambda\rightarrow \lambda_0 = \alpha^2/\omega_0$. The critical temperature is then \cite{PhysRevLett.115.247002}
\beq
T_c = \epsilon e^{-\omega_0/\alpha^2}.\label{eq:TcvanHove}
\eeq
With constant electron-phonon coupling, this means that $\omega_0 \rightarrow \omega_0(u)$. Differentiating Eq. \eqref{eq:TcvanHove} with respect to strain gives:
\beq
\frac{\partial_u T_c(u)}{T_c(u)} = -\frac{\partial_u \omega_0(u)}{\alpha^2}. 
\eeq

Experiments in stressed bulk STO found a linear dependence of $\omega^2$ on stress for both the structural and ferroelectric phonon modes \cite{PhysRevB.13.271} and we therefore consider that the soft modes at $q=0$ have a general form with applied strain: $\omega_0^2(u) = \omega_0^2(0) + b u $ \cite{PhysRevB.13.271}. Calculating $\partial_u \omega_0(u)$ gives:
\beq
\frac{\partial_u T_c(u)}{T_c(u)} = \frac{-b}{\alpha^2\omega_0(u)}.
\eeq
The immediate consequence is a sharp rise in the derivative of $T_c$ with respect to $u$ near criticality where $\omega_0 \rightarrow 0$. The divergence in the derivative is a consequence of the simple model we use and is not physical, yet the peak in $T_c$ as a result of the quantum critical point is physical and should be observable experimentally.

\begin{figure}[t!]
\includegraphics[width=\columnwidth]{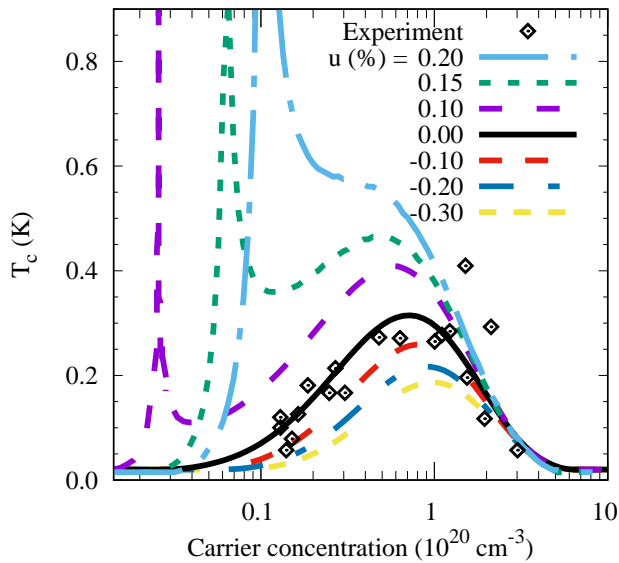} 
\caption{(Colour online) Superconducting domes for several values of uniform strain, ($-0.3\%\leq u\leq 0.2\%$) in 3D using the model of Ref. \onlinecite{PhysRevLett.115.247002} and the strain dependence of the ferroelectric mode frequencies from Density Functional Theory (DFT) calculations. The dome broadens and $T_c$ increases under dilation, with the appearance of a peak at low carrier concentrations that is a signature of $\omega_q(u)=0$ in Eq. \eqref{eq:lambdaInt}. Overall scale of maximum $T_c$ is set by $T_0 = 7 K$, used here. Experimental data from Ref. \onlinecite{PhysRev.163.380}. \label{fig:SCDomesVarStrain}}
\end{figure}

In Fig. \ref{fig:SCDomesVarStrain}, superconducting domes, constructed using the model of Ref. \onlinecite{PhysRevLett.115.247002}, and including the dependence of both the phonon spectra and Fermi energy on strain (detail in the supplemental material \S II), are plotted for several values of uniform strain in 3D. The key features of Fig. \ref{fig:SCDomesVarStrain} are a strong increase in the critical temperature under tensile strains, accompanied by a shift of the maximum $T_c$ towards lower carrier densities, a broadening of the superconducting dome and the appearance of a sharp secondary peak in $T_c(n)$ at low carrier concentrations, the value of which is limited by the range of temperatures considered and the tuning over the carrier density. Under compression, the dome is narrowed and the critical temperature decreases.

The strong peak at low carrier densities is the direct result of the softening of the ferroelectric modes explicitly present in the coupling [Eq. \eqref{eq:lambdaInt}] and the peaks in $T_c$ correspond exactly to the carrier densities where $\omega_0(u) = 0$. Its presence in the case of oxygen isotope substitution \cite{PhysRevLett.115.247002} was not observed because, in that model, it occurs at much lower carrier densities. Away from these soft mode induced peaks, the maximum $T_c$ varies linearly with strain, and a strong deviation, including, as seen at lower carrier concentrations, the possibility for the largest $T_c$ to occur at intermediate strains, would indicate a soft mode in the coupling mechanism. The strong dependence of the ferroelectric modes in STO on strain potentially allows access to interesting new features by bringing them within the range of carrier concentrations that is relevant for superconductivity, although the strong, relatively narrow peaks observed here are direct consequences of the model used.

\begin{figure}[t!]
\includegraphics[width=\columnwidth]{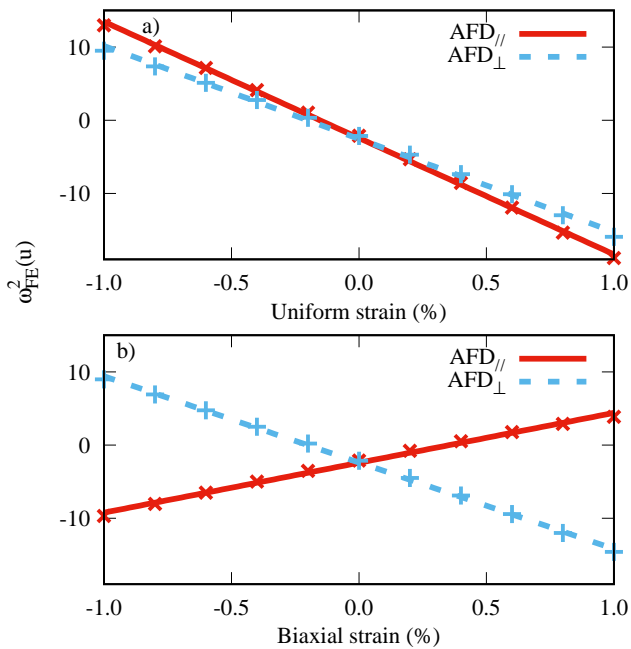}
\caption{(Colour online) Squared frequencies of the ferroelectric modes with strains calculated using DFT (details in supplemental material \S III) and their linear fits with slope $b$. a) Uniform strain: $b = -15.9, -12.7$ for modes parallel (red, solid) and perpendicular (blue, dashed) to the antiferrodistortive (AFD) axis; hardening of both modes under compression and softening under dilation characterises `asymmetric' behaviour. b) Biaxial strain in the basal plane: $b = 6.8, -11.8$, respectively; that one mode softens while the other hardens under both compression and dilation allows `symmetric' behaviour of $T_c$ whereby it increases or decreases for all strains. The crosses indicate the DFT data points. \label{fig:FitstoFreqSquared}}
\end{figure}

The value of $b=-13$ (strain as a percentage) is representative of the linear fits to the squared phonon frequencies in Fig. \ref{fig:FitstoFreqSquared}. A smaller value of $|b|$ would result in weaker changes in $T_c$ and the divergence of $T_c$ occurring at a lower carrier density for a given strain. A positive value of $b$ would lead to an increase in $T_c$ and the divergence moving towards higher carrier concentrations under compression.


We can characterise two types of general response: an `asymmetric' response occurs when the relevant ferroelectric mode softens (frequency decreases) under one sign of strain and hardens for the other so the change in $T_c$ under tensile (compressive) strain is an extension of the behaviour under compressive (tensile) strain. A `symmetric' response is characterised by the variation in $T_c$ having the same sign for all strains although the gradients may not necessarily be the same on each side of $u=0$.

As seen from the squared frequencies of the ferroelectric modes plotted in Fig. \ref{fig:FitstoFreqSquared}, all individual modes give asymmetric behaviour under strain; symmetric behaviour is possible in the case of biaxial strain if the critical temperature is controlled by the \textit{softest} mode rather than dependent on a particular mode. Therefore such symmetric or asymmetric behaviour under different types of strain could indicate whether the critical temperature is determined by the softest (lowest frequency) ferroelectric mode, or linked to a specific mode or orientation of the tetragonal c-axis, providing an important insight into the superconducting coupling mechanism of STO.


Having examined the effects of strain on the superconducting dome assuming a ferroelectric soft mode character for the superconducting coupling, we now develop a simple Ginzburg-Landau model of the strained superconducting system to capture the dominant features of the change in the critical temperature. We focus on the asymmetric response and assume that the coupling strength is independent of strain. 

The (Helmholtz) free energy has three parts: the unstrained uniform superconductor in the absence of an applied magnetic field with coefficients $\alpha = a(T-T_c^0)$ and $\beta$ \cite{RepProgPhys.36.103}; Hooke's law with strain $u$ and elastic constants $\zeta$ \cite{PhilMag.1949.Devonshire, PhysRevB.13.271, PhysRevB.61.825}; a part describing the direct coupling between strain and the uniform order parameter $\psi$ with coupling strengths $\gamma$ \cite{PhysRevLett.65.2294, Sci.344.6181.283.SI}.
\begin{eqnarray}
F &=& \alpha |\psi|^2 + \frac{\beta}{2}|\psi|^4 +\frac{\zeta_{11}}{2}u_{1}^2 + \frac{\zeta_{22}}{2}u^2_{2} + \zeta_{12}u_{1}u_{2} \nonumber \\
&& \qquad + (\gamma_{1}u_{1} + \gamma_{2}u_{2})|\psi|^2. \label{eq:FESCStrain}
\end{eqnarray}
In order to focus on the qualitative behaviours that may occur, we have restricted our analysis to a single (uniform) superconducting order parameter $\psi$ in two dimensions with strains applied along the principal axes of a square lattice. The strains $u_1$ and $u_2$ therefore denote fractional changes in length along the the [100] and [010] directions of the tetragonal unit cell of low temperature STO respectively \cite{PhysRevB.61.825}. The in-plane shear strain $u_{6}$ is assumed to be zero and the volume preserving reaction in the [001] direction is neglected.

Minimising Eq. \eqref{eq:FESCStrain} with respect to $\psi^*$ gives a linear change in the critical temperature with applied strain:
\beq
\Delta T_{c} = T_c^s - T_c^0 = - u_{1}\Gamma_{1} - u_{2}\Gamma_{2}, \label{eq:GenDelTcStrain}
\eeq
where scaled coupling constants, $\Gamma = \gamma/a$, with units of temperature (K) have been introduced. In the case of uniform, symmetry preserving strain, $u_{1}=u_{2} = u$, and we define $\Gamma = \Gamma_{1}+\Gamma_{2}$ so Eq. \eqref{eq:GenDelTcStrain} simplifies to:
\beq
\Delta T_c^{lm} = -\Gamma u \label{eq:DelTcWBxStrain}.
\eeq
There is a simple linear dependence of $T_c$ on uniform strain and the behaviour would reflect an asymmetric nature of the ferroelectric modes. Detail of how an estimate for $\Gamma$ can be extracted from experimental pressure data can be found in the supplemental material \S IV.

\begin{figure}[t!]
\includegraphics[width=\columnwidth]{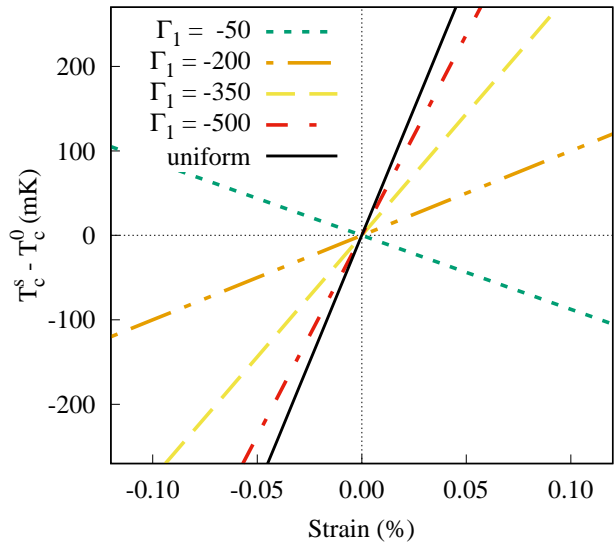}
\caption{(Colour online) Change in $T_c$ as a result of uniaxial applied strain $u_1$ with $u_2 = -\nu u_1$. Varying (negative) $\Gamma_{1}$ while keeping $\Gamma_{1}+ \Gamma_{2} = -600$K leads to both an increase and a decrease of $T_c$ under uniaxial compression depending on the exact value of $\Gamma_1$. (Uniform is $u_{1} = u_{2}$.) \label{fig:SimpleUniaxial_ArbGamma}}
\end{figure}

Meanwhile, when the only applied strain is $u_1$, the relaxation of the lattice determines $u_2$ such that so $u_{2} = -\nu u_{1}$ where $\nu$ is Poisson's ratio \cite{Lautrup_ContMatt} ($\nu \approx 0.25$ for bulk STO \cite{PhilMagA.80.3.621}), and the coupling strengths to strains in the different directions become important. For simplicity, we neglect the effects of the change of the crystal symmetry and assume that, at least for small strains, the dominant irreducible representation will be that of the unstrained system, though, in principle they will differ \cite{PhysRevB.36.5186}. The change in the critical temperature is again linear in the applied (uniaxial) strain:
\beq
\Delta T_{c}^{ux} = - u_{1}(\Gamma_{1} - \nu\Gamma_{2}).
\eeq

Assuming that $\Gamma_1$ and $\Gamma_2$ have the same sign, the relative sizes of $\Gamma_1$ and $\Gamma_2$ will determine the sign of $\Delta T_c^{ux}$, with no change in $T_c$ due to uniaxial strain if $\Gamma_1 = \nu\Gamma_2$. In Fig. \ref{fig:SimpleUniaxial_ArbGamma}, the value of $\Gamma=\Gamma_{1}+\Gamma_{2}$ is fixed at $\Gamma=-600$K as determined from hydrostatic pressure data and $\Delta T_c^{ux}$ for different choices of $\Gamma_{1}$ plotted showing that both decreases and increases in $T_c$ under uniaxial compressive strains are possible. 

In experiments, linear $\Delta T_c$ under uniaxial compression in the [100] direction has been observed to be either positive or negative depending on the carrier concentration, while in all other cases, compression lead to a decrease in $T_c$ that was strongest for hydrostatic pressure \cite{JLowTPhys.2.333}. The analysis presented here indicates that at different points across the superconducting dome the relative sizes of $\Gamma_{1}$ and $\Gamma_{2}$ may differ while their sum is (approximately) constant. Since the observed increase in $T_c$ under uniaxial compression is weak compared with the decrease under hydrostatic pressure \cite{JLowTPhys.2.333}, the possibility of $\Gamma_1 >0$ has not been considered.

In the above discussion, we considered a 2D superconducting order parameter and the results presented are qualitative in nature. Due to the preserved symmetry of the square lattice, the case of uniform strain (biaxial in 2D) reflects hydrostatic pressure while the uniaxial case is representative of all other strains that where the lattice relaxes in directions perpendicular to the applied strain but otherwise preserve the pseudocubic structure of the unit cell. The value of $\Gamma$ is overestimated, but the behaviours representative of those that may occur.

The interplay between ferroelectricity and superconductivity in STO has been examined for both oxygen isotope substitution \cite{PhysRevLett.115.247002} and a combination of oxygen depletion and calcium doping \cite{NatPhys.13.7.643}, with critical carrier densities $10^{19}<10^{20}\mathrm{cm}^{-3}$ beyond which the ferroelectric-like order is destroyed \cite{PhysRevLett.115.247002, NatPhys.13.7.643}. Both bulk \cite{PhysRevB.13.271} and thin film STO \cite{PhysRevB.61.825}, samples become ferroelectric beyond critical strains on the range of $0.3 \%$ \cite{PhysRevB.61.825, PhysRevB.71.024102, PhysRevB.73.184112, ApplPhysLett.96.232902, PhysRevB.73.184112, PhysRevB.71.024102}. Assuming that the critical carrier density that destroys ferroelectric order in STO does not depend strongly on the origin of the ferroelectric order, we expect ferroelectric order and superconductivity to occur within the same range of carrier densities for strained systems. The proposed strain tuneability of STO superconductivity builds on these ideas and provides an alternative test of the role of ferroelectric criticality in STO superconductivity with several distinct signatures. While the discussion has focussed on bulk STO, we expect similar effects to occur in interfacial superconductivity  of STO based systems.

In conclusion, we have considered the effects of strain on superconductivity in STO in two situations. First, we assumed that the superconducting coupling is caused by the ferroelectric modes that are present due to the incipient ferroelectric nature of STO. This led to an increase of $T_c$ under uniform tensile strain, a broadening of the superconducting dome - most noticeable at lower carrier densities, and a shift of the maximum $T_c$ towards lower carrier densities; with exactly opposite behaviours under compression. We note that differences between uniform and biaxial strain experiments may provide insight into the relative importance of the ferroelectric modes parallel and perpendicular to the tetragonal c-axis for the superconducting pairing. Although reference has been made to STO, the behaviours observed are expected to be general for any superconductor where pairing is mediated by softening ferroelectric modes. One important feature of strained STO is that small tensile strains are sufficient to bring the ferroelectric quantum critical point, characterised by $\omega_0(u) = 0$, to carrier densities that are well within the superconducting dome. Thus strain tuning is expected to be a versatile means of investigating the interplay between superconductivity and ferroelectricity in STO.

In order to understand experimental data of linear changes of $T_c$ under various strain configurations, we also considered a simple Ginzburg-Landau analysis of a uniform 2D superconductor under strain in which there is a linear dependence - very strong for STO - of $T_c$ on strain. The observed qualitative differences between uniaxial strain and hydrostatic pressure \cite{JLowTPhys.2.333} can be understood by considering that the couplings to strain in different directions may depend on carrier density while their sum remains fixed.

{\it Acknowledgments}. We are grateful for enlightening discussions with C. Triola and Y. Kedem. The work was supported by the US DOE BES E3B7, by VILLUM FONDEN via the Centre of Excellence for Dirac Materials (Grant No. 11744) and Knut and Alice Wallenberg Foundation, and the European Research Council under the European Union's Seventh Framework Program (FP/2207-2013)/ERC Grant Agreement No. DM-321031. A.N. and N.A.S. acknowledge support from ETH-Zurich. Calculations were performed at ETH-Zurich (Euler cluster) and at the Swiss National Supercomputing Centre (project ID p504).

%


\begin{thebibliography}{38}%
\makeatletter
\providecommand \@ifxundefined [1]{%
 \@ifx{#1\undefined}
}%
\providecommand \@ifnum [1]{%
 \ifnum #1\expandafter \@firstoftwo
 \else \expandafter \@secondoftwo
 \fi
}%
\providecommand \@ifx [1]{%
 \ifx #1\expandafter \@firstoftwo
 \else \expandafter \@secondoftwo
 \fi
}%
\providecommand \natexlab [1]{#1}%
\providecommand \enquote  [1]{``#1''}%
\providecommand \bibnamefont  [1]{#1}%
\providecommand \bibfnamefont [1]{#1}%
\providecommand \citenamefont [1]{#1}%
\providecommand \href@noop [0]{\@secondoftwo}%
\providecommand \href [0]{\begingroup \@sanitize@url \@href}%
\providecommand \@href[1]{\@@startlink{#1}\@@href}%
\providecommand \@@href[1]{\endgroup#1\@@endlink}%
\providecommand \@sanitize@url [0]{\catcode `\\12\catcode `\$12\catcode
  `\&12\catcode `\#12\catcode `\^12\catcode `\_12\catcode `\%12\relax}%
\providecommand \@@startlink[1]{}%
\providecommand \@@endlink[0]{}%
\providecommand \url  [0]{\begingroup\@sanitize@url \@url }%
\providecommand \@url [1]{\endgroup\@href {#1}{\urlprefix }}%
\providecommand \urlprefix  [0]{URL }%
\providecommand \Eprint [0]{\href }%
\providecommand \doibase [0]{http://dx.doi.org/}%
\providecommand \selectlanguage [0]{\@gobble}%
\providecommand \bibinfo  [0]{\@secondoftwo}%
\providecommand \bibfield  [0]{\@secondoftwo}%
\providecommand \translation [1]{[#1]}%
\providecommand \BibitemOpen [0]{}%
\providecommand \bibitemStop [0]{}%
\providecommand \bibitemNoStop [0]{.\EOS\space}%
\providecommand \EOS [0]{\spacefactor3000\relax}%
\providecommand \BibitemShut  [1]{\csname bibitem#1\endcsname}%
\let\auto@bib@innerbib\@empty
\bibitem [{\citenamefont {Worlock}\ and\ \citenamefont
  {Fleury}(1967)}]{PhysRevLett.19.1176}%
  \BibitemOpen
  \bibfield  {author} {\bibinfo {author} {\bibfnamefont {J.~M.}\ \bibnamefont
  {Worlock}}\ and\ \bibinfo {author} {\bibfnamefont {P.~A.}\ \bibnamefont
  {Fleury}},\ }\href {\doibase 10.1103/PhysRevLett.19.1176} {\bibfield
  {journal} {\bibinfo  {journal} {Phys. Rev. Lett.}\ }\textbf {\bibinfo
  {volume} {19}},\ \bibinfo {pages} {1176} (\bibinfo {year}
  {1967})}\BibitemShut {NoStop}%
\bibitem [{\citenamefont {Stirling}(1972)}]{JPhysCSolStat.5.2711}%
  \BibitemOpen
  \bibfield  {author} {\bibinfo {author} {\bibfnamefont {W.~G.}\ \bibnamefont
  {Stirling}},\ }
  {\bibfield  {journal} {\bibinfo  {journal} {Journal of Physics C: Solid State Physics}\ }
  \textbf {\bibinfo {volume} {5}},\ \bibinfo {pages} {2711}
  (\bibinfo {year} {1972})}\BibitemShut {NoStop}%
\bibitem [{\citenamefont {Uwe}\ and\ \citenamefont
  {Sakudo}(1976)}]{PhysRevB.13.271}%
  \BibitemOpen
  \bibfield  {author} {\bibinfo {author} {\bibfnamefont {H.}~\bibnamefont
  {Uwe}}\ and\ \bibinfo {author} {\bibfnamefont {T.}~\bibnamefont {Sakudo}},\
  }\href {\doibase 10.1103/PhysRevB.13.271} {\bibfield  {journal} {\bibinfo
  {journal} {Phys. Rev. B}\ }\textbf {\bibinfo {volume} {13}},\ \bibinfo
  {pages} {271} (\bibinfo {year} {1976})}\BibitemShut {NoStop}%
\bibitem [{\citenamefont {M\"uller}\ and\ \citenamefont
  {Burkard}(1979)}]{PhysRevB.19.3593}%
  \BibitemOpen
  \bibfield  {author} {\bibinfo {author} {\bibfnamefont {K.~A.}\ \bibnamefont
  {M\"uller}}\ and\ \bibinfo {author} {\bibfnamefont {H.}~\bibnamefont
  {Burkard}},\ }\href {\doibase 10.1103/PhysRevB.19.3593} {\bibfield  {journal}
  {\bibinfo  {journal} {Phys. Rev. B}\ }\textbf {\bibinfo {volume} {19}},\
  \bibinfo {pages} {3593} (\bibinfo {year} {1979})}\BibitemShut {NoStop}%
\bibitem [{\citenamefont {Jang}\ \emph {et~al.}(2010)\citenamefont {Jang},
  \citenamefont {Kumar}, \citenamefont {Denev}, \citenamefont {Biegalski},
  \citenamefont {Maksymovych}, \citenamefont {Bark}, \citenamefont {Nelson},
  \citenamefont {Folkman}, \citenamefont {Baek}, \citenamefont {Balke},
  \citenamefont {Brooks}, \citenamefont {Tenne}, \citenamefont {Schlom},
  \citenamefont {Chen}, \citenamefont {Pan}, \citenamefont {Kalinin},
  \citenamefont {Gopalan},\ and\ \citenamefont {Eom}}]{PhysRevLett.104.197601}%
  \BibitemOpen
  \bibfield  {author} {\bibinfo {author} {\bibfnamefont {H.~W.}\ \bibnamefont
  {Jang}}, \bibinfo {author} {\bibfnamefont {A.}~\bibnamefont {Kumar}},
  \bibinfo {author} {\bibfnamefont {S.}~\bibnamefont {Denev}}, \bibinfo
  {author} {\bibfnamefont {M.~D.}\ \bibnamefont {Biegalski}}, \bibinfo {author}
  {\bibfnamefont {P.}~\bibnamefont {Maksymovych}}, \bibinfo {author}
  {\bibfnamefont {C.~W.}\ \bibnamefont {Bark}}, \bibinfo {author}
  {\bibfnamefont {C.~T.}\ \bibnamefont {Nelson}}, \bibinfo {author}
  {\bibfnamefont {C.~M.}\ \bibnamefont {Folkman}}, \bibinfo {author}
  {\bibfnamefont {S.~H.}\ \bibnamefont {Baek}}, \bibinfo {author}
  {\bibfnamefont {N.}~\bibnamefont {Balke}}, \bibinfo {author} {\bibfnamefont
  {C.~M.}\ \bibnamefont {Brooks}}, \bibinfo {author} {\bibfnamefont {D.~A.}\
  \bibnamefont {Tenne}}, \bibinfo {author} {\bibfnamefont {D.~G.}\ \bibnamefont
  {Schlom}}, \bibinfo {author} {\bibfnamefont {L.~Q.}\ \bibnamefont {Chen}},
  \bibinfo {author} {\bibfnamefont {X.~Q.}\ \bibnamefont {Pan}}, \bibinfo
  {author} {\bibfnamefont {S.~V.}\ \bibnamefont {Kalinin}}, \bibinfo {author}
  {\bibfnamefont {V.}~\bibnamefont {Gopalan}}, \ and\ \bibinfo {author}
  {\bibfnamefont {C.~B.}\ \bibnamefont {Eom}},\ }\href {\doibase
  10.1103/PhysRevLett.104.197601} {\bibfield  {journal} {\bibinfo  {journal}
  {Phys. Rev. Lett.}\ }\textbf {\bibinfo {volume} {104}},\ \bibinfo {pages}
  {197601} (\bibinfo {year} {2010})}\BibitemShut {NoStop}%
\bibitem [{\citenamefont {Pertsev}\ \emph {et~al.}(2000)\citenamefont
  {Pertsev}, \citenamefont {Tagantsev},\ and\ \citenamefont
  {Setter}}]{PhysRevB.61.825}%
  \BibitemOpen
  \bibfield  {author} {\bibinfo {author} {\bibfnamefont {N.~A.}\ \bibnamefont
  {Pertsev}}, \bibinfo {author} {\bibfnamefont {A.~K.}\ \bibnamefont
  {Tagantsev}}, \ and\ \bibinfo {author} {\bibfnamefont {N.}~\bibnamefont
  {Setter}},\ }\href {\doibase 10.1103/PhysRevB.61.R825} {\bibfield  {journal}
  {\bibinfo  {journal} {Phys. Rev. B}\ }\textbf {\bibinfo {volume} {61}},\
  \bibinfo {pages} {R825} (\bibinfo {year} {2000})}\BibitemShut {NoStop}%
\bibitem [{\citenamefont {Burke}\ and\ \citenamefont
  {Pressley}(1971)}]{SolStatComm.9.191}%
  \BibitemOpen
  \bibfield  {author} {\bibinfo {author} {\bibfnamefont {W.~J.}\ \bibnamefont
  {Burke}}\ and\ \bibinfo {author} {\bibfnamefont {R.~J.}\ \bibnamefont
  {Pressley}},\ }\href {\doibase
  http://dx.doi.org/10.1016/0038-1098(71)90115-3} {\bibfield  {journal}
  {\bibinfo  {journal} {Solid State Communications}\ }\textbf {\bibinfo
  {volume} {9}},\ \bibinfo {pages} {191 } (\bibinfo {year} {1971})}\BibitemShut
  {NoStop}%
\bibitem [{\citenamefont {Rischau}\ \emph {et~al.}(2017)\citenamefont
  {Rischau}, \citenamefont {Lin}, \citenamefont {Grams}, \citenamefont {Finck},
  \citenamefont {Hams}, \citenamefont {Engelmayer}, \citenamefont {Lorenz},
  \citenamefont {Gallais}, \citenamefont {Fauqu\'{e}}, \citenamefont
  {Hemberger},\ and\ \citenamefont {Behnia}}]{NatPhys.13.7.643}%
  \BibitemOpen
  \bibfield  {author} {\bibinfo {author} {\bibfnamefont {C.~W.}\ \bibnamefont
  {Rischau}}, \bibinfo {author} {\bibfnamefont {X.}~\bibnamefont {Lin}},
  \bibinfo {author} {\bibfnamefont {C.~P.}\ \bibnamefont {Grams}}, \bibinfo
  {author} {\bibfnamefont {D.}~\bibnamefont {Finck}}, \bibinfo {author}
  {\bibfnamefont {S.}~\bibnamefont {Hams}}, \bibinfo {author} {\bibfnamefont
  {J.}~\bibnamefont {Engelmayer}}, \bibinfo {author} {\bibfnamefont
  {T.}~\bibnamefont {Lorenz}}, \bibinfo {author} {\bibfnamefont
  {Y.}~\bibnamefont {Gallais}}, \bibinfo {author} {\bibfnamefont
  {B.}~\bibnamefont {Fauqu\'{e}}}, \bibinfo {author} {\bibfnamefont
  {J.}~\bibnamefont {Hemberger}}, \ and\ \bibinfo {author} {\bibfnamefont
  {K.}~\bibnamefont {Behnia}},\ }\href@noop {} {\bibfield  {journal} {\bibinfo
  {journal} {Nature Physics}\ }\textbf {\bibinfo {volume} {13}},\ \bibinfo
  {pages} {643} (\bibinfo {year} {2017})}\BibitemShut {NoStop}%
\bibitem [{\citenamefont {Edge}\ \emph {et~al.}(2015)\citenamefont {Edge},
  \citenamefont {Kedem}, \citenamefont {Aschauer}, \citenamefont {Spaldin},\
  and\ \citenamefont {Balatsky}}]{PhysRevLett.115.247002}%
  \BibitemOpen
  \bibfield  {author} {\bibinfo {author} {\bibfnamefont {J.~M.}\ \bibnamefont
  {Edge}}, \bibinfo {author} {\bibfnamefont {Y.}~\bibnamefont {Kedem}},
  \bibinfo {author} {\bibfnamefont {U.}~\bibnamefont {Aschauer}}, \bibinfo
  {author} {\bibfnamefont {N.~A.}\ \bibnamefont {Spaldin}}, \ and\ \bibinfo
  {author} {\bibfnamefont {A.~V.}\ \bibnamefont {Balatsky}},\ }\href {\doibase
  10.1103/PhysRevLett.115.247002} {\bibfield  {journal} {\bibinfo  {journal}
  {Phys. Rev. Lett.}\ }\textbf {\bibinfo {volume} {115}},\ \bibinfo {pages}
  {247002} (\bibinfo {year} {2015})}\BibitemShut {NoStop}%
\bibitem [{\citenamefont {Stucky}\ \emph {et~al.}(2016)\citenamefont {Stucky},
  \citenamefont {Scheerer}, \citenamefont {Ren}, \citenamefont {Jaccard},
  \citenamefont {Poumirol}, \citenamefont {Barreteau}, \citenamefont
  {Giannini},\ and\ \citenamefont {van~der Marel}}]{SciRep.6.37582}%
  \BibitemOpen
  \bibfield  {author} {\bibinfo {author} {\bibfnamefont {A.}~\bibnamefont
  {Stucky}}, \bibinfo {author} {\bibfnamefont {G.~W.}\ \bibnamefont
  {Scheerer}}, \bibinfo {author} {\bibfnamefont {Z.}~\bibnamefont {Ren}},
  \bibinfo {author} {\bibfnamefont {D.}~\bibnamefont {Jaccard}}, \bibinfo
  {author} {\bibfnamefont {J.-M.}\ \bibnamefont {Poumirol}}, \bibinfo {author}
  {\bibfnamefont {C.}~\bibnamefont {Barreteau}}, \bibinfo {author}
  {\bibfnamefont {E.}~\bibnamefont {Giannini}}, \ and\ \bibinfo {author}
  {\bibfnamefont {D.}~\bibnamefont {van~der Marel}},\ }\href@noop {} {\bibfield
   {journal} {\bibinfo  {journal} {Scientific Reports}\ }\textbf {\bibinfo
  {volume} {6}} (\bibinfo {year} {2016})}\BibitemShut {NoStop}%
\bibitem [{\citenamefont {Pfeiffer}\ and\ \citenamefont
  {Schooley}(1970)}]{JLowTPhys.2.333}%
  \BibitemOpen
  \bibfield  {author} {\bibinfo {author} {\bibfnamefont {E.~R.}\ \bibnamefont
  {Pfeiffer}}\ and\ \bibinfo {author} {\bibfnamefont {J.~F.}\ \bibnamefont
  {Schooley}},\ }\href@noop {} {\bibfield  {journal} {\bibinfo  {journal}
  {Journal of Low Temperature Physics}\ }\textbf {\bibinfo {volume} {2}},\
  \bibinfo {pages} {333} (\bibinfo {year} {1970})}\BibitemShut {NoStop}%
\bibitem [{\citenamefont {Brandt}\ and\ \citenamefont
  {Ginzburg}(1969)}]{ContempPhys.10.4.355}%
  \BibitemOpen
  \bibfield  {author} {\bibinfo {author} {\bibfnamefont {N.~B.}\ \bibnamefont
  {Brandt}}\ and\ \bibinfo {author} {\bibfnamefont {N.~I.}\ \bibnamefont
  {Ginzburg}},\ }\href@noop {} {\bibfield  {journal} {\bibinfo  {journal}
  {Contempory Physics}\ }\textbf {\bibinfo {volume} {10}},\ \bibinfo {pages}
  {355} (\bibinfo {year} {1969})}\BibitemShut {NoStop}%
\bibitem [{\citenamefont {Koonce}\ \emph {et~al.}(1967)\citenamefont {Koonce},
  \citenamefont {Cohen}, \citenamefont {Schooley}, \citenamefont {Hosler},\
  and\ \citenamefont {Pfeiffer}}]{PhysRev.163.380}%
  \BibitemOpen
  \bibfield  {author} {\bibinfo {author} {\bibfnamefont {C.~S.}\ \bibnamefont
  {Koonce}}, \bibinfo {author} {\bibfnamefont {M.~L.}\ \bibnamefont {Cohen}},
  \bibinfo {author} {\bibfnamefont {J.~F.}\ \bibnamefont {Schooley}}, \bibinfo
  {author} {\bibfnamefont {W.~R.}\ \bibnamefont {Hosler}}, \ and\ \bibinfo
  {author} {\bibfnamefont {E.~R.}\ \bibnamefont {Pfeiffer}},\ }\href {\doibase
  10.1103/PhysRev.163.380} {\bibfield  {journal} {\bibinfo  {journal} {Phys.
  Rev.}\ }\textbf {\bibinfo {volume} {163}},\ \bibinfo {pages} {380} (\bibinfo
  {year} {1967})}\BibitemShut {NoStop}%
\bibitem [{\citenamefont {Lin}\ \emph {et~al.}(2014)\citenamefont {Lin},
  \citenamefont {Bridoux}, \citenamefont {Gourgout}, \citenamefont {Seyfarth},
  \citenamefont {Kr\"amer}, \citenamefont {Nardone}, \citenamefont {Fauqu\'e},\
  and\ \citenamefont {Behnia}}]{PhysRevLett.112.207002}%
  \BibitemOpen
  \bibfield  {author} {\bibinfo {author} {\bibfnamefont {X.}~\bibnamefont
  {Lin}}, \bibinfo {author} {\bibfnamefont {G.}~\bibnamefont {Bridoux}},
  \bibinfo {author} {\bibfnamefont {A.}~\bibnamefont {Gourgout}}, \bibinfo
  {author} {\bibfnamefont {G.}~\bibnamefont {Seyfarth}}, \bibinfo {author}
  {\bibfnamefont {S.}~\bibnamefont {Kr\"amer}}, \bibinfo {author}
  {\bibfnamefont {M.}~\bibnamefont {Nardone}}, \bibinfo {author} {\bibfnamefont
  {B.}~\bibnamefont {Fauqu\'e}}, \ and\ \bibinfo {author} {\bibfnamefont
  {K.}~\bibnamefont {Behnia}},\ }\href {\doibase
  10.1103/PhysRevLett.112.207002} {\bibfield  {journal} {\bibinfo  {journal}
  {Phys. Rev. Lett.}\ }\textbf {\bibinfo {volume} {112}},\ \bibinfo {pages}
  {207002} (\bibinfo {year} {2014})}\BibitemShut {NoStop}%
\bibitem [{\citenamefont {Thiemann}\ \emph {et~al.}(2017)\citenamefont
  {Thiemann}, \citenamefont {Beutel}, \citenamefont {Dressel}, \citenamefont
  {Lee-{H}one}, \citenamefont {Broun}, \citenamefont {Fillis-{T}sirakis},
  \citenamefont {Boschker}, \citenamefont {Mannhart},\ and\ \citenamefont
  {Scheffler}}]{ArXiv:1703.04716}%
  \BibitemOpen
  \bibfield  {author} {\bibinfo {author} {\bibfnamefont {M.}~\bibnamefont
  {Thiemann}}, \bibinfo {author} {\bibfnamefont {M.~H.}\ \bibnamefont
  {Beutel}}, \bibinfo {author} {\bibfnamefont {M.}~\bibnamefont {Dressel}},
  \bibinfo {author} {\bibfnamefont {N.~R.}\ \bibnamefont {Lee-{H}one}},
  \bibinfo {author} {\bibfnamefont {D.~M.}\ \bibnamefont {Broun}}, \bibinfo
  {author} {\bibfnamefont {E.}~\bibnamefont {Fillis-{T}sirakis}}, \bibinfo
  {author} {\bibfnamefont {H.}~\bibnamefont {Boschker}}, \bibinfo {author}
  {\bibfnamefont {J.}~\bibnamefont {Mannhart}}, \ and\ \bibinfo {author}
  {\bibfnamefont {M.}~\bibnamefont {Scheffler}},\ }\href@noop {} {\bibfield
  {journal} {\bibinfo  {journal} {arXiv preprint arXiv:1703.04716}\ } (\bibinfo
  {year} {2017})}\BibitemShut {NoStop}%
\bibitem [{\citenamefont {van~der Marel}\ \emph {et~al.}(2011)\citenamefont
  {van~der Marel}, \citenamefont {van Mechelen},\ and\ \citenamefont
  {Mazin}}]{PhysRevB.84.205111}%
  \BibitemOpen
  \bibfield  {author} {\bibinfo {author} {\bibfnamefont {D.}~\bibnamefont
  {van~der Marel}}, \bibinfo {author} {\bibfnamefont {J.~L.~M.}\ \bibnamefont
  {van Mechelen}}, \ and\ \bibinfo {author} {\bibfnamefont {I.~I.}\
  \bibnamefont {Mazin}},\ }\href {\doibase 10.1103/PhysRevB.84.205111}
  {\bibfield  {journal} {\bibinfo  {journal} {Phys. Rev. B}\ }\textbf {\bibinfo
  {volume} {84}},\ \bibinfo {pages} {205111} (\bibinfo {year}
  {2011})}\BibitemShut {NoStop}%
\bibitem [{\citenamefont {Cyrot}(1973)}]{RepProgPhys.36.103}%
  \BibitemOpen
  \bibfield  {author} {\bibinfo {author} {\bibfnamefont {M.}~\bibnamefont
  {Cyrot}},\ }\href {http://stacks.iop.org/0034-4885/36/i=2/a=001} {\bibfield
  {journal} {\bibinfo  {journal} {Reports on Progress in Physics}\ }\textbf
  {\bibinfo {volume} {36}},\ \bibinfo {pages} {103} (\bibinfo {year}
  {1973})}\BibitemShut {NoStop}%
\bibitem [{\citenamefont {Devonshire}(1949)}]{PhilMag.1949.Devonshire}%
  \BibitemOpen
  \bibfield  {author} {\bibinfo {author} {\bibfnamefont {A.~F.}\ \bibnamefont
  {Devonshire}},\ }\href@noop {} {\bibfield  {journal} {\bibinfo  {journal}
  {The London, Edinburgh, and Dublin Philosophical Magazine and Journal of
  Science}\ }\textbf {\bibinfo {volume} {40}},\ \bibinfo {pages} {1040}
  (\bibinfo {year} {1949})}\BibitemShut {NoStop}%
\bibitem [{\citenamefont {Bruls}\ \emph {et~al.}(1990)\citenamefont {Bruls},
  \citenamefont {Weber}, \citenamefont {Wolf}, \citenamefont {Thalmeier},
  \citenamefont {L\"uthi}, \citenamefont {Visser},\ and\ \citenamefont
  {Menovsky}}]{PhysRevLett.65.2294}%
  \BibitemOpen
  \bibfield  {author} {\bibinfo {author} {\bibfnamefont {G.}~\bibnamefont
  {Bruls}}, \bibinfo {author} {\bibfnamefont {D.}~\bibnamefont {Weber}},
  \bibinfo {author} {\bibfnamefont {B.}~\bibnamefont {Wolf}}, \bibinfo {author}
  {\bibfnamefont {P.}~\bibnamefont {Thalmeier}}, \bibinfo {author}
  {\bibfnamefont {B.}~\bibnamefont {L\"uthi}}, \bibinfo {author} {\bibfnamefont
  {A.~d.}\ \bibnamefont {Visser}}, \ and\ \bibinfo {author} {\bibfnamefont
  {A.}~\bibnamefont {Menovsky}},\ }\href {\doibase 10.1103/PhysRevLett.65.2294}
  {\bibfield  {journal} {\bibinfo  {journal} {Phys. Rev. Lett.}\ }\textbf
  {\bibinfo {volume} {65}},\ \bibinfo {pages} {2294} (\bibinfo {year}
  {1990})}\BibitemShut {NoStop}%
\bibitem [{\citenamefont {Hicks}\ \emph
  {et~al.}(2014{\natexlab{a}})\citenamefont {Hicks}, \citenamefont {Brodsky},
  \citenamefont {Yelland}, \citenamefont {Gibbs}, \citenamefont {Bruin},
  \citenamefont {Barber}, \citenamefont {Edkins}, \citenamefont {Nishimura},
  \citenamefont {Yonezawa}, \citenamefont {Maeno},\ and\ \citenamefont
  {Mackenzie}}]{Sci.344.6181.283.SI}%
  \BibitemOpen
  \bibfield  {author} {\bibinfo {author} {\bibfnamefont {C.~W.}\ \bibnamefont
  {Hicks}}, \bibinfo {author} {\bibfnamefont {D.~O.}\ \bibnamefont {Brodsky}},
  \bibinfo {author} {\bibfnamefont {E.~A.}\ \bibnamefont {Yelland}}, \bibinfo
  {author} {\bibfnamefont {A.~S.}\ \bibnamefont {Gibbs}}, \bibinfo {author}
  {\bibfnamefont {J.~A.~N.}\ \bibnamefont {Bruin}}, \bibinfo {author}
  {\bibfnamefont {M.~E.}\ \bibnamefont {Barber}}, \bibinfo {author}
  {\bibfnamefont {S.~D.}\ \bibnamefont {Edkins}}, \bibinfo {author}
  {\bibfnamefont {K.}~\bibnamefont {Nishimura}}, \bibinfo {author}
  {\bibfnamefont {S.}~\bibnamefont {Yonezawa}}, \bibinfo {author}
  {\bibfnamefont {Y.}~\bibnamefont {Maeno}}, \ and\ \bibinfo {author}
  {\bibfnamefont {A.~P.}\ \bibnamefont {Mackenzie}},\ }\href {\doibase
  10.1126/science.1248292} {\bibfield  {journal} {\bibinfo  {journal}
  {Science}\ }\textbf {\bibinfo {volume} {344}},\ \bibinfo {pages} {283}
  (\bibinfo {year} {2014}{\natexlab{a}})},\ \bibinfo {note} {supplementary
  material for Ref. \onlinecite{Sci.344.6181.283}}\BibitemShut {NoStop}%
\bibitem [{\citenamefont {Lautrup}(2005)}]{Lautrup_ContMatt}%
  \BibitemOpen
  \bibfield  {author} {\bibinfo {author} {\bibfnamefont {B.}~\bibnamefont
  {Lautrup}},\ }\href@noop {} {\emph {\bibinfo {title} {Physics of continuous
  matter}}}\ (\bibinfo  {publisher} {Io{P} publishing},\ \bibinfo {year}
  {2005})\BibitemShut {NoStop}%
\bibitem [{\citenamefont {Suzuki}\ \emph {et~al.}(2000)\citenamefont {Suzuki},
  \citenamefont {Nishi},\ and\ \citenamefont {Fujimoto}}]{PhilMagA.80.3.621}%
  \BibitemOpen
  \bibfield  {author} {\bibinfo {author} {\bibfnamefont {T.}~\bibnamefont
  {Suzuki}}, \bibinfo {author} {\bibfnamefont {Y.}~\bibnamefont {Nishi}}, \
  and\ \bibinfo {author} {\bibfnamefont {M.}~\bibnamefont {Fujimoto}},\
  }\href@noop {} {\bibfield  {journal} {\bibinfo  {journal} {Philosophical
  Magazine A}\ }\textbf {\bibinfo {volume} {80}},\ \bibinfo {pages} {621}
  (\bibinfo {year} {2000})}\BibitemShut {NoStop}%
\bibitem [{\citenamefont {Sigrist}\ \emph {et~al.}(1987)\citenamefont
  {Sigrist}, \citenamefont {Joynt},\ and\ \citenamefont
  {Rice}}]{PhysRevB.36.5186}%
  \BibitemOpen
  \bibfield  {author} {\bibinfo {author} {\bibfnamefont {M.}~\bibnamefont
  {Sigrist}}, \bibinfo {author} {\bibfnamefont {R.}~\bibnamefont {Joynt}}, \
  and\ \bibinfo {author} {\bibfnamefont {T.~M.}\ \bibnamefont {Rice}},\ }\href
  {\doibase 10.1103/PhysRevB.36.5186} {\bibfield  {journal} {\bibinfo
  {journal} {Phys. Rev. B}\ }\textbf {\bibinfo {volume} {36}},\ \bibinfo
  {pages} {5186} (\bibinfo {year} {1987})}\BibitemShut {NoStop}%
\bibitem [{\citenamefont {Antons}\ \emph {et~al.}(2005)\citenamefont {Antons},
  \citenamefont {Neaton}, \citenamefont {Rabe},\ and\ \citenamefont
  {Vanderbilt}}]{PhysRevB.71.024102}%
  \BibitemOpen
  \bibfield  {author} {\bibinfo {author} {\bibfnamefont {A.}~\bibnamefont
  {Antons}}, \bibinfo {author} {\bibfnamefont {J.~B.}\ \bibnamefont {Neaton}},
  \bibinfo {author} {\bibfnamefont {K.~M.}\ \bibnamefont {Rabe}}, \ and\
  \bibinfo {author} {\bibfnamefont {D.}~\bibnamefont {Vanderbilt}},\ }\href
  {\doibase 10.1103/PhysRevB.71.024102} {\bibfield  {journal} {\bibinfo
  {journal} {Phys. Rev. B}\ }\textbf {\bibinfo {volume} {71}},\ \bibinfo
  {pages} {024102} (\bibinfo {year} {2005})}\BibitemShut {NoStop}%
\bibitem [{\citenamefont {Li}\ \emph {et~al.}(2006)\citenamefont {Li},
  \citenamefont {Choudhury}, \citenamefont {Haeni}, \citenamefont {Biegalski},
  \citenamefont {Vasudevarao}, \citenamefont {Sharan}, \citenamefont {Ma},
  \citenamefont {Levy}, \citenamefont {Gopalan}, \citenamefont
  {Trolier-McKinstry}, \citenamefont {Schlom}, \citenamefont {Jia},\ and\
  \citenamefont {Chen}}]{PhysRevB.73.184112}%
  \BibitemOpen
  \bibfield  {author} {\bibinfo {author} {\bibfnamefont {Y.~L.}\ \bibnamefont
  {Li}}, \bibinfo {author} {\bibfnamefont {S.}~\bibnamefont {Choudhury}},
  \bibinfo {author} {\bibfnamefont {J.~H.}\ \bibnamefont {Haeni}}, \bibinfo
  {author} {\bibfnamefont {M.~D.}\ \bibnamefont {Biegalski}}, \bibinfo {author}
  {\bibfnamefont {A.}~\bibnamefont {Vasudevarao}}, \bibinfo {author}
  {\bibfnamefont {A.}~\bibnamefont {Sharan}}, \bibinfo {author} {\bibfnamefont
  {H.~Z.}\ \bibnamefont {Ma}}, \bibinfo {author} {\bibfnamefont
  {J.}~\bibnamefont {Levy}}, \bibinfo {author} {\bibfnamefont {V.}~\bibnamefont
  {Gopalan}}, \bibinfo {author} {\bibfnamefont {S.}~\bibnamefont
  {Trolier-McKinstry}}, \bibinfo {author} {\bibfnamefont {D.~G.}\ \bibnamefont
  {Schlom}}, \bibinfo {author} {\bibfnamefont {Q.~X.}\ \bibnamefont {Jia}}, \
  and\ \bibinfo {author} {\bibfnamefont {L.~Q.}\ \bibnamefont {Chen}},\ }\href
  {\doibase 10.1103/PhysRevB.73.184112} {\bibfield  {journal} {\bibinfo
  {journal} {Phys. Rev. B}\ }\textbf {\bibinfo {volume} {73}},\ \bibinfo
  {pages} {184112} (\bibinfo {year} {2006})}\BibitemShut {NoStop}%
\bibitem [{\citenamefont {Sheng}\ \emph {et~al.}(2010)\citenamefont {Sheng},
  \citenamefont {Li}, \citenamefont {Zhang}, \citenamefont {Choudhury},
  \citenamefont {Jia}, \citenamefont {Gopalan}, \citenamefont {Schlom},
  \citenamefont {Liu},\ and\ \citenamefont {Chen}}]{ApplPhysLett.96.232902}%
  \BibitemOpen
  \bibfield  {author} {\bibinfo {author} {\bibfnamefont {G.}~\bibnamefont
  {Sheng}}, \bibinfo {author} {\bibfnamefont {Y.~L.}\ \bibnamefont {Li}},
  \bibinfo {author} {\bibfnamefont {J.~X.}\ \bibnamefont {Zhang}}, \bibinfo
  {author} {\bibfnamefont {S.}~\bibnamefont {Choudhury}}, \bibinfo {author}
  {\bibfnamefont {Q.~X.}\ \bibnamefont {Jia}}, \bibinfo {author} {\bibfnamefont
  {V.}~\bibnamefont {Gopalan}}, \bibinfo {author} {\bibfnamefont {D.~G.}\
  \bibnamefont {Schlom}}, \bibinfo {author} {\bibfnamefont {Z.~K.}\
  \bibnamefont {Liu}}, \ and\ \bibinfo {author} {\bibfnamefont {L.~Q.}\
  \bibnamefont {Chen}},\ }\href@noop {} {\bibfield  {journal} {\bibinfo
  {journal} {Applied Physics Letters}\ }\textbf {\bibinfo {volume} {96}},\
  \bibinfo {pages} {232902} (\bibinfo {year} {2010})}\BibitemShut {NoStop}%
\bibitem [{\citenamefont {Hicks}\ \emph
  {et~al.}(2014{\natexlab{b}})\citenamefont {Hicks}, \citenamefont {Brodsky},
  \citenamefont {Yelland}, \citenamefont {Gibbs}, \citenamefont {Bruin},
  \citenamefont {Barber}, \citenamefont {Edkins}, \citenamefont {Nishimura},
  \citenamefont {Yonezawa}, \citenamefont {Maeno},\ and\ \citenamefont
  {Mackenzie}}]{Sci.344.6181.283}%
  \BibitemOpen
  \bibfield  {author} {\bibinfo {author} {\bibfnamefont {C.~W.}\ \bibnamefont
  {Hicks}}, \bibinfo {author} {\bibfnamefont {D.~O.}\ \bibnamefont {Brodsky}},
  \bibinfo {author} {\bibfnamefont {E.~A.}\ \bibnamefont {Yelland}}, \bibinfo
  {author} {\bibfnamefont {A.~S.}\ \bibnamefont {Gibbs}}, \bibinfo {author}
  {\bibfnamefont {J.~A.~N.}\ \bibnamefont {Bruin}}, \bibinfo {author}
  {\bibfnamefont {M.~E.}\ \bibnamefont {Barber}}, \bibinfo {author}
  {\bibfnamefont {S.~D.}\ \bibnamefont {Edkins}}, \bibinfo {author}
  {\bibfnamefont {K.}~\bibnamefont {Nishimura}}, \bibinfo {author}
  {\bibfnamefont {S.}~\bibnamefont {Yonezawa}}, \bibinfo {author}
  {\bibfnamefont {Y.}~\bibnamefont {Maeno}}, \ and\ \bibinfo {author}
  {\bibfnamefont {A.~P.}\ \bibnamefont {Mackenzie}},\ }\href {\doibase
  10.1126/science.1248292} {\bibfield  {journal} {\bibinfo  {journal}
  {Science}\ }\textbf {\bibinfo {volume} {344}},\ \bibinfo {pages} {283}
  (\bibinfo {year} {2014}{\natexlab{b}})}\BibitemShut {NoStop}%
\bibitem [{\citenamefont {Kresse}\ and\ \citenamefont
  {Furthm{\"u}ller}(1996)}]{kresse1996efficient}%
  \BibitemOpen
  \bibfield  {author} {\bibinfo {author} {\bibfnamefont {G.}~\bibnamefont
  {Kresse}}\ and\ \bibinfo {author} {\bibfnamefont {J.}~\bibnamefont
  {Furthm{\"u}ller}},\ }\href@noop {} {\bibfield  {journal} {\bibinfo
  {journal} {Phys. Rev. B}\ }\textbf {\bibinfo {volume} {54}},\ \bibinfo
  {pages} {11169} (\bibinfo {year} {1996})}\BibitemShut {NoStop}%
\bibitem [{\citenamefont {Perdew}\ \emph {et~al.}(2008)\citenamefont {Perdew},
  \citenamefont {Ruzsinszky}, \citenamefont {Csonka}, \citenamefont {Vydrov},
  \citenamefont {Scuseria}, \citenamefont {Constantin}, \citenamefont {Zhou},\
  and\ \citenamefont {Burke}}]{perdew2008restoring}%
  \BibitemOpen
  \bibfield  {author} {\bibinfo {author} {\bibfnamefont {J.~P.}\ \bibnamefont
  {Perdew}}, \bibinfo {author} {\bibfnamefont {A.}~\bibnamefont {Ruzsinszky}},
  \bibinfo {author} {\bibfnamefont {G.~I.}\ \bibnamefont {Csonka}}, \bibinfo
  {author} {\bibfnamefont {O.~A.}\ \bibnamefont {Vydrov}}, \bibinfo {author}
  {\bibfnamefont {G.~E.}\ \bibnamefont {Scuseria}}, \bibinfo {author}
  {\bibfnamefont {L.~A.}\ \bibnamefont {Constantin}}, \bibinfo {author}
  {\bibfnamefont {X.}~\bibnamefont {Zhou}}, \ and\ \bibinfo {author}
  {\bibfnamefont {K.}~\bibnamefont {Burke}},\ }\href@noop {} {\bibfield
  {journal} {\bibinfo  {journal} {Physical Review Letters}\ }\textbf {\bibinfo
  {volume} {100}},\ \bibinfo {pages} {136406} (\bibinfo {year}
  {2008})}\BibitemShut {NoStop}%
\bibitem [{\citenamefont {Togo}\ and\ \citenamefont
  {Tanaka}(2015)}]{togo2015first}%
  \BibitemOpen
  \bibfield  {author} {\bibinfo {author} {\bibfnamefont {A.}~\bibnamefont
  {Togo}}\ and\ \bibinfo {author} {\bibfnamefont {I.}~\bibnamefont {Tanaka}},\
  }\href@noop {} {\bibfield  {journal} {\bibinfo  {journal} {Scr. Mater.}\
  }\textbf {\bibinfo {volume} {108}},\ \bibinfo {pages} {1} (\bibinfo {year}
  {2015})}\BibitemShut {NoStop}%
\bibitem [{\citenamefont {Okazaki}\ and\ \citenamefont
  {Kawaminami}(1973)}]{MatResBull.8.545}%
  \BibitemOpen
  \bibfield  {author} {\bibinfo {author} {\bibfnamefont {A.}~\bibnamefont
  {Okazaki}}\ and\ \bibinfo {author} {\bibfnamefont {M.}~\bibnamefont
  {Kawaminami}},\ }\href {\doibase
  http://dx.doi.org/10.1016/0025-5408(73)90130-X} {\bibfield  {journal}
  {\bibinfo  {journal} {Materials Research Bulletin}\ }\textbf {\bibinfo
  {volume} {8}},\ \bibinfo {pages} {545} (\bibinfo {year} {1973})}\BibitemShut
  {NoStop}%
\bibitem [{\citenamefont {Lytle}(1964)}]{JApplPhys.35.2212}%
  \BibitemOpen
  \bibfield  {author} {\bibinfo {author} {\bibfnamefont {F.~W.}\ \bibnamefont
  {Lytle}},\ }\href@noop {} {\bibfield  {journal} {\bibinfo  {journal} {Journal
  of Applied Physics}\ }\textbf {\bibinfo {volume} {35}},\ \bibinfo {pages}
  {2212} (\bibinfo {year} {1964})}\BibitemShut {NoStop}%
\bibitem [{\citenamefont {Rimai}\ and\ \citenamefont
  {deMars}(1962)}]{PhysRev.127.702}%
  \BibitemOpen
  \bibfield  {author} {\bibinfo {author} {\bibfnamefont {L.}~\bibnamefont
  {Rimai}}\ and\ \bibinfo {author} {\bibfnamefont {G.~A.}\ \bibnamefont
  {deMars}},\ }\href {\doibase 10.1103/PhysRev.127.702} {\bibfield  {journal}
  {\bibinfo  {journal} {Phys. Rev.}\ }\textbf {\bibinfo {volume} {127}},\
  \bibinfo {pages} {702} (\bibinfo {year} {1962})}\BibitemShut {NoStop}%
\bibitem [{\citenamefont {Fleury}\ \emph {et~al.}(1968)\citenamefont {Fleury},
  \citenamefont {Scott},\ and\ \citenamefont {Worlock}}]{PhysRevLett.21.16}%
  \BibitemOpen
  \bibfield  {author} {\bibinfo {author} {\bibfnamefont {P.~A.}\ \bibnamefont
  {Fleury}}, \bibinfo {author} {\bibfnamefont {J.~F.}\ \bibnamefont {Scott}}, \
  and\ \bibinfo {author} {\bibfnamefont {J.~M.}\ \bibnamefont {Worlock}},\
  }\href {\doibase 10.1103/PhysRevLett.21.16} {\bibfield  {journal} {\bibinfo
  {journal} {Phys. Rev. Lett.}\ }\textbf {\bibinfo {volume} {21}},\ \bibinfo
  {pages} {16} (\bibinfo {year} {1968})}\BibitemShut {NoStop}%
\bibitem [{\citenamefont {Bell}\ and\ \citenamefont
  {Rupprecht}(1963)}]{PhysRev.129.90}%
  \BibitemOpen
  \bibfield  {author} {\bibinfo {author} {\bibfnamefont {R.~O.}\ \bibnamefont
  {Bell}}\ and\ \bibinfo {author} {\bibfnamefont {G.}~\bibnamefont
  {Rupprecht}},\ }\href {\doibase 10.1103/PhysRev.129.90} {\bibfield  {journal}
  {\bibinfo  {journal} {Phys. Rev.}\ }\textbf {\bibinfo {volume} {129}},\
  \bibinfo {pages} {90} (\bibinfo {year} {1963})}\BibitemShut {NoStop}%
\bibitem [{\citenamefont {Rupprecht}\ and\ \citenamefont
  {Winter}(1967)}]{PhysRev.155.1019}%
  \BibitemOpen
  \bibfield  {author} {\bibinfo {author} {\bibfnamefont {G.}~\bibnamefont
  {Rupprecht}}\ and\ \bibinfo {author} {\bibfnamefont {W.~H.}\ \bibnamefont
  {Winter}},\ }\href {\doibase 10.1103/PhysRev.155.1019} {\bibfield  {journal}
  {\bibinfo  {journal} {Phys. Rev.}\ }\textbf {\bibinfo {volume} {155}},\
  \bibinfo {pages} {1019} (\bibinfo {year} {1967})}\BibitemShut {NoStop}%
\bibitem [{\citenamefont {Bark}\ \emph {et~al.}(2011)\citenamefont {Bark},
  \citenamefont {Felker}, \citenamefont {Wang}, \citenamefont {Zhang},
  \citenamefont {Jang}, \citenamefont {Folkman}, \citenamefont {Park},
  \citenamefont {Baek}, \citenamefont {Zhou}, \citenamefont {Fong},
  \citenamefont {Pan}, \citenamefont {Tsymbal}, \citenamefont {Rzchowski},\
  and\ \citenamefont {Eom}}]{PNAS.108.12.4720}%
  \BibitemOpen
  \bibfield  {author} {\bibinfo {author} {\bibfnamefont {C.~W.}\ \bibnamefont
  {Bark}}, \bibinfo {author} {\bibfnamefont {D.~A.}\ \bibnamefont {Felker}},
  \bibinfo {author} {\bibfnamefont {Y.}~\bibnamefont {Wang}}, \bibinfo {author}
  {\bibfnamefont {Y.}~\bibnamefont {Zhang}}, \bibinfo {author} {\bibfnamefont
  {H.~W.}\ \bibnamefont {Jang}}, \bibinfo {author} {\bibfnamefont {C.~M.}\
  \bibnamefont {Folkman}}, \bibinfo {author} {\bibfnamefont {J.~W.}\
  \bibnamefont {Park}}, \bibinfo {author} {\bibfnamefont {S.~H.}\ \bibnamefont
  {Baek}}, \bibinfo {author} {\bibfnamefont {H.}~\bibnamefont {Zhou}}, \bibinfo
  {author} {\bibfnamefont {D.~D.}\ \bibnamefont {Fong}}, \bibinfo {author}
  {\bibfnamefont {X.~Q.}\ \bibnamefont {Pan}}, \bibinfo {author} {\bibfnamefont
  {E.~Y.}\ \bibnamefont {Tsymbal}}, \bibinfo {author} {\bibfnamefont {M.~S.}\
  \bibnamefont {Rzchowski}}, \ and\ \bibinfo {author} {\bibfnamefont {C.~B.}\
  \bibnamefont {Eom}},\ }\href@noop {} {\bibfield  {journal} {\bibinfo
  {journal} {Proceedings of the National Academy of Sciences}\ }\textbf
  {\bibinfo {volume} {108}},\ \bibinfo {pages} {4720} (\bibinfo {year}
  {2011})}\BibitemShut {NoStop}%
\bibitem [{\citenamefont {Noad}\ \emph {et~al.}(2016)\citenamefont {Noad},
  \citenamefont {Spanton}, \citenamefont {Nowack}, \citenamefont {Inoue},
  \citenamefont {Kim}, \citenamefont {Merz}, \citenamefont {Bell},
  \citenamefont {Hikita}, \citenamefont {Xu}, \citenamefont {Liu},
  \citenamefont {Vailionis}, \citenamefont {Hwang},\ and\ \citenamefont
  {Moler}}]{PhysRevB.94.174516}%
  \BibitemOpen
  \bibfield  {author} {\bibinfo {author} {\bibfnamefont {H.}~\bibnamefont
  {Noad}}, \bibinfo {author} {\bibfnamefont {E.~M.}\ \bibnamefont {Spanton}},
  \bibinfo {author} {\bibfnamefont {K.~C.}\ \bibnamefont {Nowack}}, \bibinfo
  {author} {\bibfnamefont {H.}~\bibnamefont {Inoue}}, \bibinfo {author}
  {\bibfnamefont {M.}~\bibnamefont {Kim}}, \bibinfo {author} {\bibfnamefont
  {T.~A.}\ \bibnamefont {Merz}}, \bibinfo {author} {\bibfnamefont
  {C.}~\bibnamefont {Bell}}, \bibinfo {author} {\bibfnamefont {Y.}~\bibnamefont
  {Hikita}}, \bibinfo {author} {\bibfnamefont {R.}~\bibnamefont {Xu}}, \bibinfo
  {author} {\bibfnamefont {W.}~\bibnamefont {Liu}}, \bibinfo {author}
  {\bibfnamefont {A.}~\bibnamefont {Vailionis}}, \bibinfo {author}
  {\bibfnamefont {H.~Y.}\ \bibnamefont {Hwang}}, \ and\ \bibinfo {author}
  {\bibfnamefont {K.~A.}\ \bibnamefont {Moler}},\ }\href {\doibase
  10.1103/PhysRevB.94.174516} {\bibfield  {journal} {\bibinfo  {journal} {Phys.
  Rev. B}\ }\textbf {\bibinfo {volume} {94}},\ \bibinfo {pages} {174516}
  (\bibinfo {year} {2016})}\BibitemShut {NoStop}%
\end{thebibliography}

\newpage
\pagebreak
\appendix 
\section{SUPPLEMENTAL MATERIAL \label{sec:SI}}

\subsection{I. On assuming $n(u) \approx n(0)$ \label{sec:nconst}}

The carrier density is defined as the number of free electrons $n_e$ per unit cell of volume $V$. In the unstrained tetragonal unit cell, $V= V_0 = a_0^2 c$. If strain is applied uniaxially along the [100] axis and the lattice is allowed to relax in the [010] and [001] directions, the Poisson's ratios for the relaxation are needed. Since STO has a tetragonal unit cell, they are, in principal, different.

The volume of the strained unit cell is: $V(u) = (a_0 + \delta a_{[100]})(a_0 + \delta a_{[010]})(c + \delta c_{[001]})$. The strains resulting from the controlled deformation in the [100] direction are $u_{[010]} = \delta a_{[010]}/a_0 = -\nu_a u_{[100]}$ and $u_{[001]} = \delta c_{[001]}/c_0 = -\nu_c u_{[100]}$. Thus, $V(u) = V_0(1+u_{[100]})(1-\nu_a u_{[100]})(1-\nu_c u_{[100]})$. If strains are small, then on expanding this out only the linear term is kept: $V(u) \approx V_0[1+u_{[100]}(1-\nu_a-\nu_c)]$ from which 
\beq
n(u) \approx n_0[1+u_{[100]} (1-\nu_a-\nu_c)].
\eeq
For many materials $\nu \sim 0.3$ \cite{Lautrup_ContMatt}
and for cubic STO $\nu \approx 0.25$ \cite{PhilMagA.80.3.621}. Further, the absolute maximum for $\nu$ is $0.5$ \cite{Lautrup_ContMatt} so tetragonal STO probably has $\nu_{a}\neq \nu_c \sim 0.25$, implying that 
\beqs
n(u) \sim n_0(1+0.5 u_{[100]}).
\eeqs
Uniform biaxial strain without relaxation of the tetragonal c-axis has $\nu_a=-1$ and $\nu_c=0$, in which case,
\beqs
n(u) \approx n_0 (1-2u).
\eeqs
Strain provides a small correction to $n$, with a coefficient of at most 2 (3 for 3D uniform strain where $\nu_c = -1$ too).

\subsection{II. Description of superconducting dome \label{sec:ConstructSCDome}}

The superconducting dome is constructed by combining Eliashberg strong coupling theory with the standard expression for the superconducting critical temperature \cite{PhysRevLett.115.247002, PhysRevB.84.205111}. The coupling is given by Eq. \eqref{eq:lambdaInt} of the main text \cite{PhysRevLett.115.247002}:
\beqs
\lambda = \int_0^\infty d\omega \frac{\alpha^2(\omega)}{\omega}F(\omega)
\eeqs
with $\omega=\omega_q$ and the frequencies of the soft ferroelectric mode excitations around the paraelectric ground state are given by
\beq
\omega_q^2(u) = 4\Gamma_{\!f}[\Gamma_{\!f} - 2J\cos(q)] + b u, \label{eq:omsqwithu}
\eeq
in which $\Gamma_{\!f} = A + BE_{\! f}^2 +CE_{\! f}$ with $E_{\! f}$ weakly dependent on strain through the carrier density $n(u) = n_0/([1+u]^3)\approx n_0(1-3u)$, so, for uniform strain
\beq
\lambda \sim \int_{-\pi}^\pi \frac{dq}{2\Gamma_{\!f}\sqrt{1-2J\cos(q)/\Gamma_{\!f}}}. \label{eq:lambdaJcos}
\eeq
The ratio $\Gamma_{\!f}/2J$ is unity on the ferroelectric quantum critical line so $2J=1$ is used for simplicity. The values of the parameters chosen are such that $\Gamma_{\!f}=1$ at zero doping and strain: $A = 1.14, B = 10^{-6}$K$^{-2}$, $C=2.5\times 10^{-3}$K$^{-1}$ and $D=190\mathrm{K}^{1/2}$, and $E_f$ is converted to carrier concentration to plot the superconducting dome \cite{PhysRevLett.115.247002}.

The critical temperature is \cite{PhysRevLett.115.247002, PhysRevB.84.205111}:
\beqs
1=\frac{\lambda}{2\pi^2}\int_{-E_{\! f}}^0d\epsilon N(\epsilon)\frac{\tanh(\epsilon/2T_c)}{\epsilon}
\eeqs
where $\epsilon$ is the energy relative to the Fermi energy $E_f$ and $N(\epsilon)$ is the density of states. The two limits are set by $N(\epsilon)=0$ for $\epsilon \leq -E_{\! f}$ and $\epsilon=0$ at $E_{\! f}$. For low doping, the relevant energy range is near $-E_{\! f}$ and the density of states in 3D is $N(\epsilon) \sim \sqrt{\epsilon+E_{\! f}}$. A change of variables $x=\epsilon/T_c$ is made and we have to solve \cite{PhysRevLett.115.247002, PhysRevB.84.205111}
\beq
\frac{D}{\lambda} = \sqrt{T_c}\int_{-E_{\! f}/T_c}^0 d x \sqrt{x+E_{\! f}/T_c} \frac{\tanh(x/2)}{x} 
\eeq
numerically with $\lambda$ given by Eq. \eqref{eq:lambdaJcos}.

\subsection{III. Ferroelectric mode in strained STO \label{sec:symmetricFE}}

Density functional calculations were performed using the Vienna Ab-initio Simulation Package ({\sc vasp})~\cite{kresse1996efficient}, with the PBEsol approximation to the exchange correlation functional~\cite{perdew2008restoring}. We used the default projector augmented wave pseudopotentials, and the wavefunction was expanded in plane waves up to a cutoff of $550$eV. The Brillouin zone was sampled using an $8\times 8\times 6$ $k$-point grid. We used the low temperature tetragonal structure of SrTiO$_{3}$ and relaxed the structures until the forces were less than 10$^{-4}$ eV/\AA{}. The phonon calculations were performed using the {\sc phonopy} code~\cite{togo2015first}, employing 80 atom supercells and a $4\times 4\times 6$ $k$-point mesh.

\begin{figure}[t!]
\includegraphics[width=0.9\columnwidth]{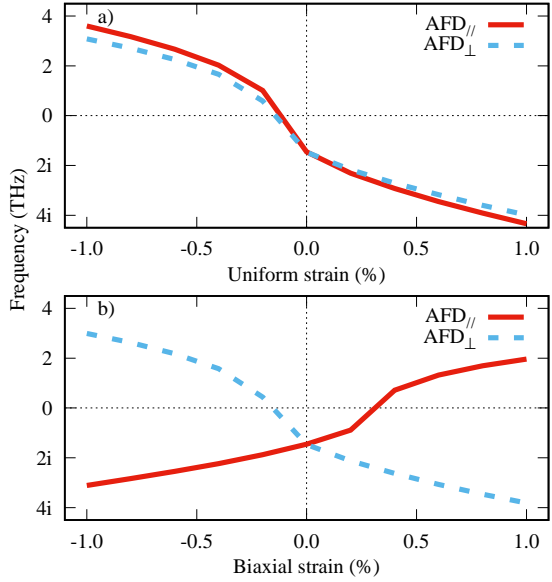}
\caption{(Colour online) Variation of the ferroelectric mode frequencies with (a) uniform change in lattice constants and (b) biaxial strain in the $ab$ plane. The blue curves are for the mode perpendicular to the axis of the antiferrodistortive (AFD) rotations and the red curves correspond to the mode parallel to the AFD axis. The imaginary frequencies correspond to unstable phonon modes. \label{Fig:fe_mode_frequencies}}
\end{figure}

We considered two scenarios: uniform change in lattice constants (corresponding to hydrostatic pressure conditions) and an $ab$ plane biaxial strain. Our results for the frequencies of the ferroelectric (FE) mode are shown in Fig.~\ref{Fig:fe_mode_frequencies}. A uniform reduction in the lattice constants hardens the FE mode and the frequency becomes positive for $\sim 0.1$\% reduction in the lattice constants [Fig.~\ref{Fig:fe_mode_frequencies}(a)]. On the other hand, with an increase in volume (negative hydrostatic pressure), the FE mode frequencies become more imaginary indicative of a stronger FE instability.

Meanwhile, under biaxial strain, the behaviour of FE modes parallel and perpendicular to the axis of antiferrodistortive (AFD) rotations, i.e. the $c$-axis, is opposite [Fig.~\ref{Fig:fe_mode_frequencies}(b)]. For an in-plane compressive strain the FE modes perpendicular to the AFD axis harden, while the mode parallel to the AFD axis becomes more unstable. Under tensile strain, the situation is reversed: modes perpendicular to the AFD axis soften and the mode parallel to the AFD axis is stabilised. Although different FE modes are softening, the overall behaviour is symmetric under strain. Uniaxial strain is expected to mirror the biaxial case analysed here, although the anisotropy between the two AFD modes may be greater and, in some geometries, the two modes perpendicular to the AFD axis may split.

\subsection{IV. Coupling strengths from pressure (stress) data \label{sec:Gibbs}}

The strain-order parameter coupling strengths must be determined from experimental data which is often in terms of applied pressure. By introducing the Gibbs free energy, $G = F -\sum_\lambda \sigma_\lambda u_\lambda$ the stresses (negative pressures) $\sigma_\lambda$ are now present. Assuming uniform strains from hydrostatic pressure, $u = u_{1} = u_{2}$; $\sigma_{1} = \sigma_{2} = \sigma \Rightarrow G = F - 2 \sigma u$, and solving $\partial G/\partial u = 0$ \cite{PhilMag.1949.Devonshire, PhysRevB.13.271, PhysRevB.61.825} gives the equilibrium value of strain:
\beqs
\overline{u} = \frac{\sigma}{\zeta} - \frac{\gamma}{2\zeta}|\psi|^2
\eeqs
which is substituted into the free energy ($\zeta = \zeta_{11}+\zeta_{12}; \gamma = \gamma_{1}+\gamma_{2}$)
\beqs
G(\sigma) = |\psi|^2\left(\alpha + \frac{\sigma\gamma}{2\zeta} \right) + \frac{|\psi|^4}{2}\left(\beta - \frac{\gamma^2}{2\zeta} \right) - \frac{\sigma^2}{\zeta}.
\eeqs
\begin{figure}[t!]
\includegraphics[width=\columnwidth]{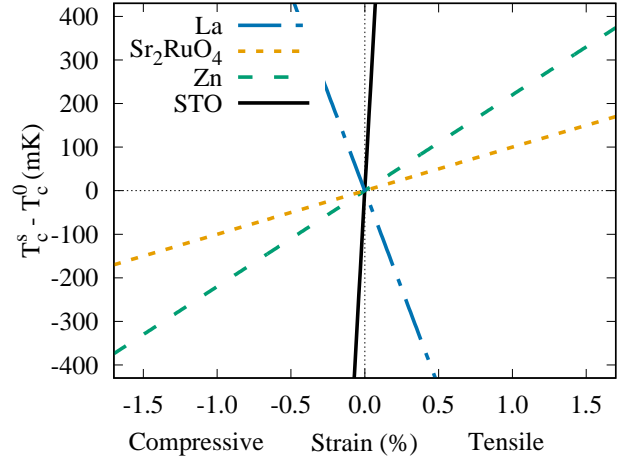}
\caption{(Colour online) Linear change in the superconducting critical temperature (asymmetric response continued from experimental behaviour under compression \cite{JLowTPhys.2.333}) as a result of uniform strain [Eq. \eqref{eq:DelTcWBxStrain} of main text] for several materials and $|u|<1.7\%$. The materials and scaled coupling strengths are: STO, $\Gamma=-600$K; Zn, $\Gamma=-20$K; Sr$_2$RuO$_4$, $\Gamma=-10$K and La, $\Gamma=90$K. \label{fig:TcwithStrain}}
\end{figure}
Minimisation with respect to $\psi^*$ gives
\beq
\Delta T_c(\sigma) = T_{c}(\sigma) - T_{c}^0 = -\frac{\Gamma\sigma}{2\zeta}. \label{eq:DelTcWStress}
\eeq

Although superconductivity in STO occurs at temperatures well below the cubic-tetragonal structural phase transition at about $105$K \cite{PhysRevB.13.271, MatResBull.8.545, JApplPhys.35.2212, PhysRev.127.702, PhysRevLett.21.16}, the elastic constants of STO at low temperatures are not well known \cite{JLowTPhys.2.333, PhysRev.129.90, PhysRev.155.1019} so we use values extrapolated from the high temperature cubic unit cell: $\zeta_{11} = \zeta_{22} = 3.36,\, \zeta_{12} = 1.07 \times 10^{11}$Pa \cite{PhysRevB.61.825, PhysRevB.13.271}.

In Fig. \ref{fig:TcwithStrain} the changes in the critical temperature that would occur with coupling constants calculated from hydrostatic pressure data for STO \cite{JLowTPhys.2.333}, zinc and lanthanum \cite{ContempPhys.10.4.355} and [110] strain data for Sr$_2$RuO$_4$ \cite{Sci.344.6181.283} (assuming [110] $\rightarrow u_1 = u_2$) are plotted for a range of strains that can be achieved in STO films by lattice mismatch to a substrate \cite{PNAS.108.12.4720}. The range of temperatures is chosen to lie near the maximum $T_c$ observed in unstrained STO \cite{PhysRev.163.380}. 

In STO, linear changes of $T_c$ under various stress configurations have been observed over a broad range ($- 0.12 \lesssim \Delta T_c \lesssim 0.031$ K) of temperatures for $T_c^0 = 0.27$ K \cite{JLowTPhys.2.333}, so, although the limit of a Landau analysis is usually $(T-T_c^0)/T_c^0 \ll 1$, the linear variation of $T_c$ obtained here is likely to be representative of strains giving a broad range of $T_c/T_c^0$. Further, changes in $T_c$ larger than $10\%$ for lattice changes of $\sim 0.1\%$ have also been observed in two dimensionally doped STO where the size of the effect was attributed to the variation of dielectric properties with the orientation of tetragonal domains  \cite{PhysRevB.94.174516}. 

The value of $\Gamma$ is certainly over estimated, as might be expected since the analysis considers a 2D system but has used hydrostatic pressure data. However, although the actual coupling strengths are not accurate, a more precise analysis would have a similar effect on all examples and the relative sizes of the $\Gamma$ values for the different materials are representative.

\end{document}